\documentclass[prb,twocolumn,superscriptaddress, showpacs]{revtex4-1}

\usepackage[dvips]{graphicx}
\graphicspath{{imagesprb/}}
\usepackage{dcolumn}
\usepackage{bm}
\usepackage{color}
\usepackage{amssymb,amsmath}

\begin{document}


\title{Spin correlations and Dzyaloshinskii-Moriya interaction in Cs$_2$CuCl$_4$}

\author{M.~A.~Fayzullin}
\affiliation{Institute for Physics, Kazan (Volga region) Federal University, 430008 Kazan, Russia}
\author{R.~M.~Eremina}
\affiliation{E.~K.~Zavoisky Physical-Technical Institute RAS, 420029 Kazan, Russia}
\affiliation{Institute for Physics, Kazan (Volga region) Federal University, 430008 Kazan, Russia}
\author{M.~V.~Eremin}
\affiliation{Institute for Physics, Kazan (Volga region) Federal University, 430008 Kazan, Russia}
\author{ A.~Dittl}
\affiliation{Experimentalphysik V, Center for Electronic Correlations and Magnetism, Institute for Physics, Augsburg University, D-86135 Augsburg, Germany}
\author{ N.~van ~Well}
\affiliation{Physikalisches Institut, Goethe-Universit\"{a}t Frankfurt, Max-von-Laue-Strasse 1, 60438 Frankfurt am Main, Germany}
\author{F.~Ritter}
\affiliation{Physikalisches Institut, Goethe-Universit\"{a}t Frankfurt, Max-von-Laue-Strasse 1, 60438 Frankfurt am Main, Germany}
\author{W.~Assmus}
\affiliation{Physikalisches Institut, Goethe-Universit\"{a}t Frankfurt, Max-von-Laue-Strasse 1, 60438 Frankfurt am Main, Germany}
\author{J.~Deisenhofer}
\affiliation{Experimentalphysik V, Center for Electronic Correlations and Magnetism, Institute for Physics, Augsburg University, D-86135 Augsburg, Germany}
\author{H.-A.~Krug von Nidda}
\author{A.~Loidl}
\affiliation{Experimentalphysik V, Center for Electronic Correlations and Magnetism, Institute for Physics, Augsburg University, D-86135 Augsburg, Germany}

\date{\today}

\begin{abstract}

We report on electron spin resonance (ESR) studies of the spin relaxation in Cs$_2$CuCl$_4$.
The main source of the ESR linewidth at temperatures $T \leq 150$~K is attributed to the uniform Dzyaloshinskii-Moriya interaction. The vector components of the Dzyaloshinskii-Moriya interaction are determined from the angular dependence of the ESR spectra using a high-temperature approximation. Both the angular and temperature dependence of the ESR linewidth have been analyzed using a self-consistent quantum-mechanical approach. In addition analytical expressions based on
a quasi-classical picture for spin fluctuations are derived, which show good agreement with the quantum-approach for temperatures $T \geq 2J/k_{\rm B} \approx 15$~K. A small modulation of the ESR linewidth observed in the $ac$-plane is attributed to the anisotropic Zeeman interaction, which reflects the two magnetically nonequivalent Cu positions.

\end{abstract}

\maketitle

\section{Introduction}
Low dimensional magnets are fascinating testing grounds for the fundamental understanding of quantum physics.
A particularly rich example of a frustrated low dimensional quantum antiferromagnet is the spin-1/2 system Cs$_2$CuCl$_4$. This material possesses an orthorhombic symmetry (space group \emph{Pnma}) with lattice
parameters $a$ = 9.753, $b$ = 7.609, and $c$ = 12.394 {\AA}.\cite{Mellor1939, Helmholz1952, Krueger2010} The spatial arrangement of the CuCl$_4$ tetrahedra leads to the formation of
antiferromagnetically coupled chains along the $b$-axis with a dominant exchange coupling constant $J = 0.62$~meV.\cite{Coldea2001,Coldea2002,Vachon2011} The exchange coupling $J^\prime = 0.117$~meV\cite{Coldea2002} between the
chains gives rise to a triangular magnetic lattice structure in the $bc$-plane. The low-temperature state below $T_{\rm N} = 0.62$\,K corresponds to a spiral order\cite{Coldea2001} and the ratio $f=|\Theta_{\rm CW}|/T_{\rm N}
\simeq$ 5 of the antiferromagnetic Curie-Weiss temperature $\Theta_{\rm CW} = -3.5$\,K\cite{Cong2011} and N\'{e}el temperature $T_{\rm N}$ signals a considerable degree of frustration in this compound. This combination of low-dimensional magnetism
and frustrated geometry has made this compound a model system to study non-trivial spin correlations and quasi-particle excitations down to low temperatures.\cite{Smirnov2012,Povarov2011,Starykh2010}

Magnetic resonance experiments have been performed previously,\cite{Sharnoff1965, Schrama1998} but more recently the low-temperature electron spin resonance (ESR) spectra observed in the spin-liquid regime have been ascribed to result from
the spinon-continuum by the influence of a uniform Dzyaloshinsky-Moriya (DM) interaction within the spin chains.\cite{Povarov2011,Smirnov2012}

Here we study the anisotropy and temperature dependent properties by ESR measurements in the paramagnetic regime and compare the microscopic spin dynamics at different temperatures. Despite the existence of a quasi two-dimensional magnetic structure, we can describe the spin-spin relaxation in terms of a linear spin chain in the presence of a uniform DM interaction. To model the ESR linewidth the corresponding two- and four-spin correlation functions are calculated in terms of a quantum approach and compared to the results of a quasi-classical approach. Moreover we study the effect of the magnetically nonequivalent Cu sites on the anisotropy of resonance field and linewidth.

\section{Experimental Details}
High quality single crystals of Cs$_2$CuCl$_4$ were grown from aqueous solution by an evaporation technique as described in Ref.~\onlinecite{Krueger2010} and characterized by magnetization and ultrasound measurements. \cite{Cong2011, Kreisel2011} ESR measurements were performed in a Bruker ELEXSYS E500 CW-spectrometer at X-band ($\nu \approx  9.36$~GHz) and Q-band frequencies ($\nu \approx  34$~GHz) equipped with continuous He-gas-flow cryostats (Oxford Instruments) in the temperature region $4 \leq T \leq 300$~K. ESR spectra display the
power $P$ absorbed by the sample from the transverse magnetic microwave field as a function of the static magnetic field $H$. The signal-to-noise ratio of the spectra is improved by recording
the derivative $dP/dH$ using lock-in technique with field modulation.

\section{Results}
\begin{figure}
\includegraphics[width=7.5cm]{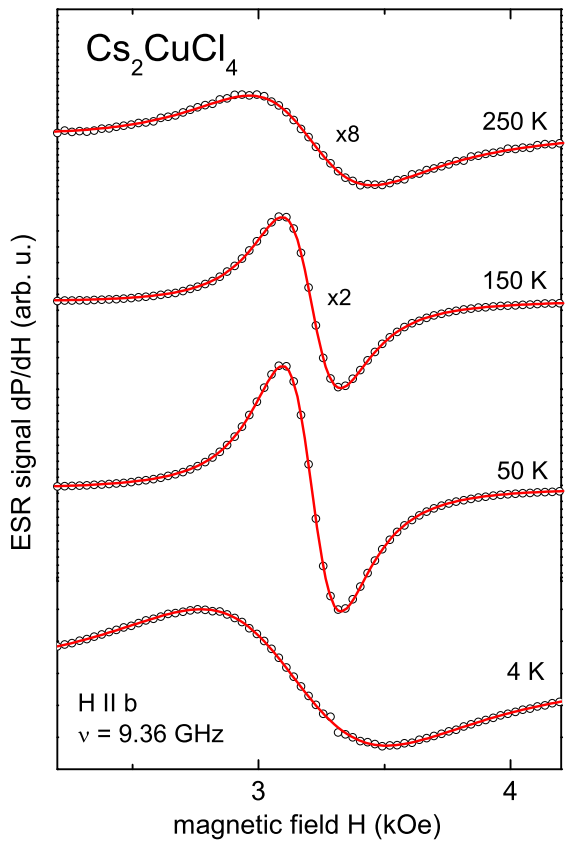}
\caption{(Color online) ESR spectra of Cs$_2$CuCl$_4$ at selected temperatures for the magnetic field applied along the crystallographic $b$-axis. The red solid lines indicate the fit by a Lorentzian line shape.} \label{spektrax0}
\end{figure}
\begin{figure}
\includegraphics[width=7.5cm]{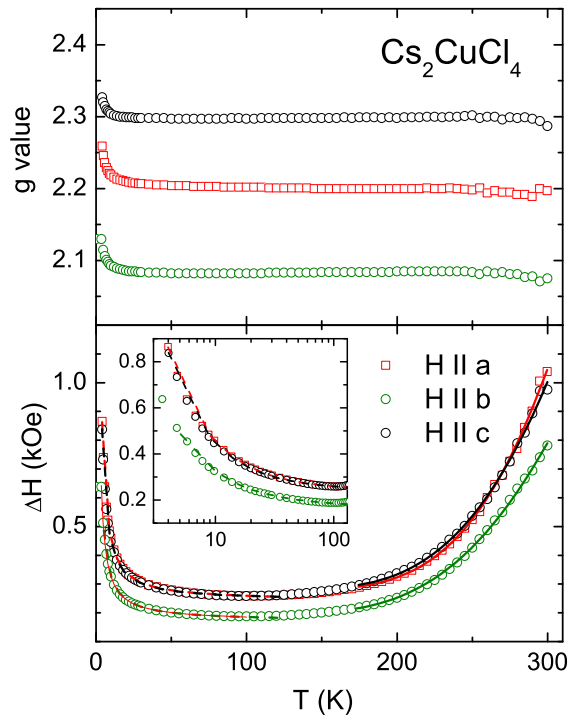}
\caption{(Color online) Temperature dependence of $g$ value (upper frame) and linewidth $\Delta H$ (lower frame) of Cs$_2$CuCl$_4$ for the magnetic field applied along the three crystallographic axes. The solid lines indicate a fit by an Arrhenius law, the dashed lines represent an empirical description in terms of a critical divergence as described in the text. The inset enlarges the low-temperature regime on a logarithmic temperature scale.}
\label{tempx0}
\end{figure}

In the whole temperature range and for all orientations of the magnetic field $H$ the ESR signal of Cs$_2$CuCl$_4$ consists of a single exchange-narrowed resonance line as exemplarily shown in
Fig.~\ref{spektrax0} for $H \parallel b$. The line is well fitted by a Lorentz shape at resonance field $H_{\rm res}$ with linewidth HWHM (half width at half maximum) $\Delta H$, indicating that spin-diffusion effects are not relevant in Cs$_2$CuCl$_4$.\cite{Dietz1971,Hennessey1973} The $g$ tensor
is obtained from resonance field and microwave frequency $\nu$ via the Larmor condition $h\nu=g\mu_{\rm B}H_{\rm res}$. It turns out to be practically independent on temperature with the
principal values $g_a = 2.20$, $g_b=2.08$, and $g_c=2.30$ and increases only slightly below 25~K as depicted in the upper frame of Fig.~\ref{tempx0}. This slight increase is most probably already related to the opening of the antiferromagnetic resonance gap to lower temperatures as reported by Povarov \textit{et al}.\cite{Povarov2011}
At the same time the linewidth exhibits three temperature regimes with different behavior as one can see in the lower frame of Fig~\ref{tempx0}. At all orientations of the magnetic field the
linewidth is only weakly temperature dependent between 50~K and 150~K. It increases following an Arrhenius law $\Delta H \propto \exp{(-\Delta/k_{\rm B}T)}$ with $\Delta/k_{\rm B} = 1600 \pm 200$~K for high temperatures
$T>150$~K. To low temperatures, for $T<50$~K the linewidth diverges on approaching magnetic order as $\Delta H \propto (T-T_{\rm N})^{-p}$ with $p = 1.2 \pm 0.1$.

Concerning the high-temperature behavior, a similar activated process has been observed for other low-dimensional magnets with Jahn-Teller active ions and spin $S = 1/2$.\cite{Heinrich2003, Zakharov2006, Eremin2008, Wang2011, Deisenhofer2012} The activation energy $\Delta/k_{\rm B} = 1600 \pm 200$~K is larger than the vibrational eigenfrequencies and, therefore, an Orbach process can be excluded. However, an energy barrier of similar size has been reported for the systems Sr$_2$VO$_4$\cite{Deisenhofer2012} and CuSb$_2$O$_6$.\cite{Heinrich2003} For the latter compound it has been proposed that the relaxation might take place via an excited state of a competing Jahn-Teller distortion,\cite{Heinrich2003} because there the activated behavior appears at temperatures just below a structural phase transition from a monoclinic into a tetragonal high-temperature phase. This phase transition of CuSb$_2$O$_6$ at about 400~K can be understood as a thermally induced averaging process between two possible elongation directions of the CuO$_6$ octahedra. But, such a structural phase transition has been observed neither in Sr$_2$VO$_4$ nor in Cs$_2$CuCl$_4$ so far. Thus, more systematic investigations are necessary to clarify this high-temperature behavior. However, a detailed analysis of such spin-lattice type relaxation phenomena is beyond the scope of the present study, where we focus on spin-spin relaxation mechanisms in the low- and intermediate temperature range.

Turning to the critical divergence at low temperatures we have to remind that, although this exponent is in the range of the expected values for three-dimensional magnets close to magnetic order,\cite{Benner1990} it is obtained from a fit of a wide temperature range above $T_{\rm N} = 0.62$~K and should be interpreted with great care, because the validity of critical exponents is usually restricted to a narrow temperature regime close to $T_{\rm N}$. Below we will show that the observed temperature dependence of the linewidth can be described by modeling a one-dimensional spin-1/2 chain in the presence of a uniform Dzyaloshinskii-Moriya (DM) interaction.

To reach this goal, we first will focus on the intermediate regime, where $J \ll k_{\rm B}T \ll \Delta$, i.e. in this temperature range the linewidth can be treated in the high-temperature limit of the theory of exchange narrowing,\cite{Anderson1953, Kubo1954} where the linewidth is broadened by pure spin-spin relaxation, but not yet by additional thermally activated processes. We will illustrate that in this temperature range the angular dependence of the linewidth gives precise information on the DM interaction.

\section{Theoretical Consideration}

If the paramagnetic resonance line is of Lorentz shape, the half width at half maximum $\Delta H$ can be calculated following the theory of exchange narrowed resonance spectra\cite{Castner1953} as
\begin{equation}
\Delta H = C\left[ \frac{M_{2}^{3}}{M_{4}} \right]^{1/2}. \label{linewidth}
\end{equation}
where $C$ is a dimensionless constant, depending on how the wings of the Lorentzian profile drop at fields of the order of the exchange field $J/g\mu_{\rm B} \gg \Delta H$.
In the simple case of a cutoff of the Lorentzian profile one obtains $C=\pi/2\sqrt{3}$. As discussed by Castner et al.,\cite{Castner1953} it may be more useful to consider a profile with exponential or gaussian wings,
which yield $C$ to be $\pi/\sqrt{2}$ or $\sqrt{\pi/2}$, respectively.
Here, $M_{2}$ and $M_{4}$ respectively denote the second moment
\begin{equation}
M_2 = \frac{\langle [{\cal H}_{\rm an},S^+][S^-,{\cal H}_{\rm an}]\rangle}{h^2\langle S^+S^-\rangle} \label{M2}
\end{equation}
and fourth moment
\begin{equation}
M_4 = \frac{\langle [{\cal H}_{\rm ex},[{\cal H}_{\rm an},S^+]][[S^-,{\cal H}_{\rm an}],{\cal H}_{\rm ex}] \rangle}{h^4\langle S^+S^-\rangle}. \label{M4}
\end{equation}
In the general case the anisotropic Hamiltonian ${\cal H}_{\rm an}$ contains the anisotropic exchange interactions, dipole-dipole interaction, and anisotropic Zeeman term. ${\cal H}_{\rm ex}$
is the isotropic exchange Hamiltonian. $S^+$ and $S^-$ denote the left and right circular components of the total spin summed up over the whole sample, respectively. As it has been shown for other Cu based spin-chain compounds like CuGeO$_3$,\cite{Pilawa1997,Eremina2003} LiCuVO$_4$,\cite{KvN} or KCuF$_3$,\cite{Eremin2008, Yamada1989} the anisotropic exchange interactions are usually dominant as compared to the dipole-dipole interaction.

A similar situation is realized in Cs$_2$CuCl$_4$ as well. Following conventional estimations,\cite{Yamada1996} the contribution to the linewidth related to the dipole-dipole interaction is about $\Delta H_{\rm DD} \sim 1$~Oe only, because the distance between nearest neighbor Cu-ions is quite large (6.3~{\AA}).\cite{Krueger2010}
An interesting specific feature of Cs$_2$CuCl$_4$, which was not pointed out in previous studies, is that in the unit cell there are two magnetically nonequivalent positions on neighboring chains. Therefore, we additionally investigated the contribution to the linewidth due to the anisotropic Zeeman interaction.
The detailed theoretical analysis is described in Appendix A. Moreover, the presence of this additional broadening mechanism of the ESR line is confirmed by measurements at higher frequency (i.e. Q-band, 34~GHz). At X-band frequency (9.36 GHz) its contribution is relatively small and thus negligible.

Therefore, we turn again to the anisotropic exchange contributions. Indeed, the uniform DM interaction has already been shown to play a major role for the properties of Cs$_2$CuCl$_4$ at low temperatures (below 4.2~K)\cite{Smirnov2012,Povarov2011} and, thus, can be expected to provide the main source of the ESR line broadening. Note that the case of Cs$_2$CuCl$_4$ significantly differs from systems like Cu benzoate,\cite{Nojiri2006} KCuGaF$_6$,\cite{Umegaki2009} and CuSe$_2$O$_5$,\cite{Herak2011} where the nonequivalent sites are alternating within the chain and the DM interaction within the chains is staggered as well. Moreover, the influence of staggered fields within the chain on the $g$-value has been discussed earlier by Nagata\cite{Nagata1976a} and is expected to increase the anisotropy of the $g$ values to low temperatures, which is not observed in Cs$_2$CuCl$_4$ (cf. Fig.~\ref{tempx0}) again corroborating the general difference of these spin systems.
%

To derive the contribution of the uniform DM interaction to the linewidth in Cs$_2$CuCl$_4$, we consider the one dimensional Heisenberg Hamiltonian
\begin{equation}
{\cal H} = \sum_i J \mathbf{S}_i\cdot\mathbf{S}_{i+1} + \sum_i \mathbf{D}\cdot[\mathbf{S}_i \times \mathbf{S}_{i+1}]. \label{Hamiltonian}
\end{equation}
where $J$ is the isotropic superexchange coupling parameter and $\bf{D}$ denotes the DM vector. In Cs$_2$CuCl$_4$ there are four different chains with different orientations of the corresponding DM vector $\bf{D}$. For their description we shall follow the notations of Starykh \textit{et al.}.\cite{Starykh2010} The calculations of the second and fourth moments were performed in a laboratory coordinate system $xyz$ with the $z$-axis chosen along the externally applied magnetic field. In order to apply the usual expressions for the transformation of the $\bf{D}$ components between laboratory $(xyz)$ and crystallographic $(XYZ)$ coordinate systems the following notations were used: $Y$ axis is chosen along the chain (i.e. crystallographic $b$-axis),  $Z$- and $X$-axis are parallel to $c$ and $a$, respectively. Then the components of the DM vector transform following
\begin{eqnarray}
D^x &=& D^X\cos{\beta}\cos{\alpha}+D^Y\cos{\beta}\sin{\alpha}-D^Z\sin{\beta}, \nonumber \\
D^y &=& D^Y\cos{\alpha}-D^X\sin{\alpha}, \label{Dtransform} \\
D^z &=& D^X\sin{\beta}\cos{\alpha}+D^Y\sin{\beta}\sin{\alpha}+D^Z\cos{\beta}, \nonumber
\end{eqnarray}
with
\begin{eqnarray}
\cos\alpha&=&\frac{A}{\sqrt{A^2+B^2}}, \cos\beta=\frac{C}{\sqrt{A^2+B^2+C^2}},\nonumber \\
g&=&\sqrt{A^2+B^2+C^2},\nonumber \\
A&=&g_{aa}\sin\theta\cos\phi+g_{ab}\sin\theta\sin\phi+g_{ac}\cos\theta, \nonumber\\
B&=&g_{ba}\sin\theta\cos\phi+g_{bb}\sin\theta\sin\phi+g_{bc}\cos\theta, \label{Dtransform1} \\
C&=&g_{ca}\sin\theta\cos\phi+g_{cb}\sin\theta\sin\phi+g_{cc}\cos\theta. \nonumber
\end{eqnarray}
Here, the angles $\alpha$ and $\beta$  define the orientation of the local coordinate system in which the Hamiltonian of the Zeeman energy takes diagonal form $g\mu_{B}H$,
while $\theta$ is the polar angle between external magnetic field and $c$ axis, and $\phi$ is the azimuthal angle counted from the $a$ direction.
Note that the orientation of these axes is essentially different for the two magnetically nonequivalent positions in the unit cell, as described in Appendix A.
These transformations should be applied for all four chains using the corresponding settings $D_1^X=D_a$, $D_1^Z=-D_c$ in chain 1, $D_2^X=-D_a$, $D_2^Z=-D_c$ in chain 2,
$D_3^X=-D_a$, $D_3^Z=D_c$ in chain 3, and $D_4^X=D_a$, $D_4^Z=D_c$ in chain 4, following the notation of Ref.~\onlinecite{Starykh2010}.

To derive the second and fourth moment for the general form of the DM interaction, we treat the spins ($S=1/2$) in terms of Pauli matrices using the corresponding commutation rules. This is in
contrast to the previous approach by Yamada and Nagata,\cite{Yamada1996,Nagata1972,Nagata1976} who treated the spins as  classical vectors. Doing so we obtain
\begin{equation}
M_2 = \frac{N D^2(\alpha,\beta)}{h^2\langle S^+S^-\rangle} \left[ \frac{1}{8}-\frac{C_1}{2}- C_2 \right] \label{M2a}
\end{equation}
for the second moment and
\begin{equation}
M_4 = \frac{N J^2 D^2(\alpha,\beta)}{h^4\langle S^+S^-\rangle} \left[ \frac{1}{16}-\frac{C_1+C_2+C_3}{4}+ C_{(4)} \right] \label{M4a}
\end{equation}
for the fourth moment with the angular dependent DM parameter
\begin{eqnarray}
& D^2(\alpha,\beta)= D_{a}^2(1+\sin^2\beta\cos^2\alpha)+D_{c}^2(1+\cos^2\beta) \nonumber \\
&+D_{a}D_{c}\cos\alpha\sin2\beta \label{D2}
\end{eqnarray}
and the pair-spin correlation functions
\begin{equation}
C_{n} = \langle S_i^{\mu} S_{i\pm n}^{\mu} \rangle, \hspace{0.5cm} (\mu=x,y,z) \label{C12}
\end{equation}
for the nearest neighbors ($n=1$), next-nearest ($n=2$) and higher order neighbors within the chain, $N$ is the total number of spins.

\begin{figure}
\includegraphics[width=7.5cm]{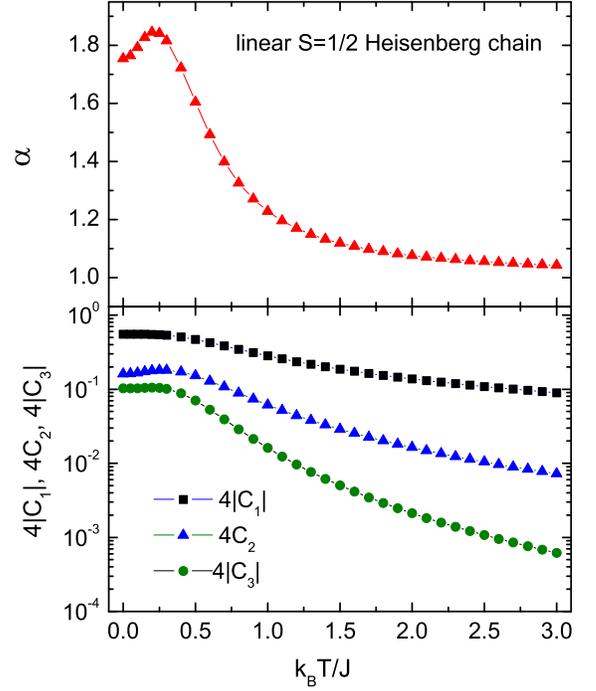}
\caption{(Color online) Calculated temperature dependence of the pair-spin correlation functions $C_n$ (lower frame) and the decoupling parameter $\alpha$, using the quantum approach as discussed in the text. The lines are drawn to guide the eyes.}\label{parameters}
\label{parameters1}
\end{figure}

The additional term $C_{(4)}$ in Eq.~\ref{M4a} is related to the four-spin correlation functions, which are usually not taken into account for simplicity and are assumed to be small. Nevertheless, the contribution of the four-spin correlation functions is not negligible in the fourth moment. In the present work we have calculated them in a quasi classical picture for spin fluctuations introduced by Fisher.\cite{Fisher1964} The details are described in Appendix B. The dominating contributions to both moments $M_2$ and $M_4$, of course, originate from the pair-spin correlation functions $C_n$. The thermodynamic average $\langle S^+S^- \rangle$, entering into the denominators of  $M_2$ and $M_4$ also can be expressed via pair-spin correlation functions as
\begin{equation}
\frac{\langle S^+ S^- \rangle}{N}=\frac{k_{\rm B}T}{J} \frac{8|C_1|}{1-12 \alpha C_1 + 4 \alpha C_2}, \label{S+S-}
\end{equation}
where $\alpha$ denotes the decoupling parameter introduced by Kondo and Yamaji\cite{Kondo1972} for the chain of Green's functions appearing in the calculations. It is determined by the "sum-rule" condition that the thermodynamic value of the spin per one site has to be $1/2$. According to the approach of Kondo and Yamaji, all spin-spin correlation functions and the decoupling parameter $\alpha$ are calculated self-consistently using the fluctuation-dissipation theorem, i.e. via the dynamic susceptibility. Below we shall call this procedure the quantum approach. A part of our results is displayed in Fig.~\ref{parameters}.

For $T > 0.5 J/k_{\rm B}$ all $C_n$ decrease monotonously on increasing temperature, and for $T \gg J/k_{\rm B}$ one finds $C_{n+1} \approx 0.1 C_{n}$, i.e. $C_1$ and $C_2$ are dominant. Only at low temperatures $T < 0.5
J/k_{\rm B}$, where the pair-spin correlations become approximately temperature independent, higher pair-spin correlations gain influence. The same holds for the decoupling parameter $\alpha$,
which essentially differs from 1 below $T < J/k_{\rm B}$.

Substituting Eq.~\ref{M2a} and Eq.~\ref{M4a} into Eq.~\ref{linewidth} we obtain the temperature and angular dependence of the  linewidth in Cs$_2$CuCl$_4$ as
\begin{equation}
\Delta H ({\rm Oe}) = C\frac{D^2(\alpha,\beta) k_{\rm B}}{Jg\mu_{\rm B}}\frac{1}{Z}\sqrt{\frac{A^3}{B}} \label{DH}
\end{equation}
with the substitutions
\begin{eqnarray}
A &=& \frac{1}{8}-\frac{C_1}{2}- C_2, \\
B &=& \frac{1}{16}-\frac{C_1+C_2+C_3}{4}+ C_{(4)}, \\
Z &=& \frac{\langle S^+ S^- \rangle}{N}.
\end{eqnarray}
We note that this expressions are valid for both approaches, quantum and quasi classical (see Appendix B). Below we will compare both variants in the discussion of the temperature dependence of the linewidth. The quantum mechanical approach, performed by a numerical self-consistent procedure as described above and the simpler quasi-classical one, described in Appendix B.

\section{Discussion and Conclusions}
\begin{figure}
\includegraphics[width=7.5cm]{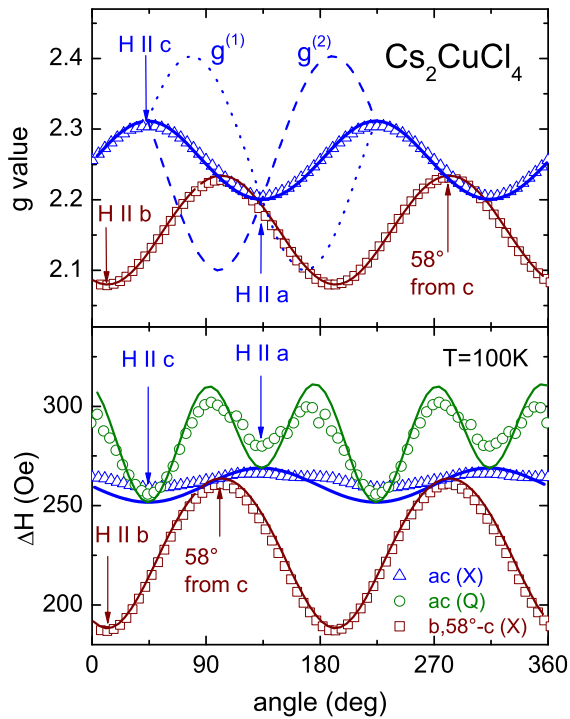}
\caption{(Color online) Angular dependence of the linewidth $\Delta H$ (lower frame) and g-value (upper frame) of Cs$_2$CuCl$_4$ at $T = 100$~K. Solid lines indicate the fitcurves by Eqs. 6 and 12, dashed and dotted lines represent the anisotropy of the $g$ tensors of the nonequivalent Cu sites in the $ac$ plane.} \label{Rotation100K}
\end{figure}

The anisotropic parameters $D_a$ and $D_c$ can be found from the angular dependence of the linewidth in the high-temperature limit of Eq.~\ref{DH} which reads
\begin{equation}
\Delta H^{T\rightarrow\infty}({\rm Oe}) = C\frac{D^2(\theta,\phi)k_{\rm B}}{Jg\mu_{\rm B}} \frac{1}{2\sqrt{2}}\label{DHinf}
\end{equation}
In Eqs.~\ref{DH} and \ref{DHinf} the parameters $D$ and $J$ are taken in Kelvin.

Figure \ref{Rotation100K} shows the angular dependence of the $g$ value and linewidth $\Delta H$ at $T=100$~K for the magnetic field applied within in the $ac$ plane as well as in a plane perpendicular to
the $ac$ plane, i.e. including the $b$ axis. This results from the fact that samples with well defined orientation could be best prepared as plates cut perpendicular to the chain ($b$) axis, or
as thin rods grown along the $b$ axis. From the coincidence of both linewidth and $g$ value one recognizes that the angular dependent data including the $b$ axis intersect the $ac$ plane at an angle of $58^{\circ}$ with respect to the $c$ axis.

The simultaneous fitting of both X-band and Q-band data sets using the infinite-temperature approximation $(T \rightarrow \infty)$ (which is applicable because of $T=100$~K $ \gg J/k_{\rm B}=7.2$~K) and taking into account the anisotropy of the $g$-factors for nonequivalent copper positions (see Appendix B)
yields $D_a=0.33$~K (6.9~GHz) and $D_c=0.36$~K (7.5~GHz) for $C=\pi/\sqrt{2}$. Povarov et al.\cite{Povarov2011} estimated values of $D_a$ and $D_c$ from the field dependence of resonance frequency as
$D_a=5.1\pm1.2$~GHz and $D_c=7.0\pm1.2$~GHz. As one can see, our results are close to the previous estimation on the assumption of an exponential dropping of the Lorentzian profile. However, it is important to note that our angular dependence analysis provides more precise information on the relative value, i.e. $D_a/D_c \approx 0.92 \pm 0.05$, of the DM components in Cs$_2$CuCl$_4$. The slight deviation of the fit curve from the X-band data in the $ac$ plane results from our simplified approach which
neglects symmetric anisotropic exchange (SAE) and anisotropic Zeeman interaction as further sources of line broadening. The latter one becomes sizeable for the Q-band data where it convincingly explains the additional $90^{\circ}$ modulation (see Appendix A), but is only of the order of a few Oe at X-band frequency. Concerning SAE we note that usually it is comparable to the dipole-dipole interaction, which, as pointed out before, is very small. Therefore, it is natural that we do not find a sizable effect of SAE in Cs$_2$CuCl$_4$.
\begin{figure}
\includegraphics[width=7.5cm]{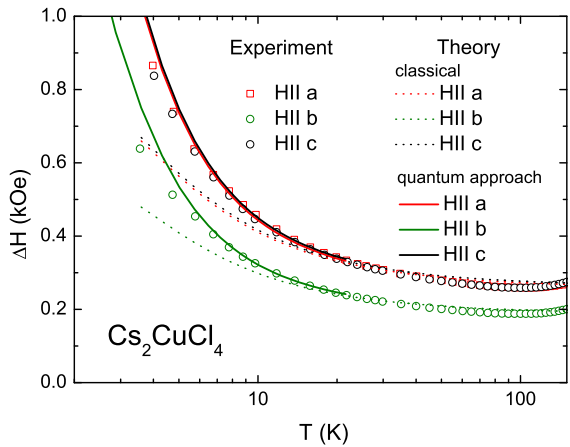}
\caption{(Color online) Calculated temperature dependence of the linewidth $\Delta H$ (Eq.~\ref{DH}) of Cs$_2$CuCl$_4$ using quantum (solid lines) and classical (dotted lines) spin correlation functions. Symbols correspond to experimental data.} \label{dHLog10}
\end{figure}

Finally, the temperature dependence of the linewidth data below 120~K again corroborates that the DM interaction is indeed the dominant broadening mechanism, as shown in Fig.~\ref{dHLog10}.
The data are satisfactorily approximated using the parameters obtained from the fitting of the angular dependence in Fig.~\ref{Rotation100K} as asymptotic values for the linewidth at $T \rightarrow \infty$. Moreover, we compare the quantum approach used in our treatment to the previous classical approach.
It is remarkable that both approaches for the temperature dependence of the spin-spin correlation functions are consistent to each other and yield results in agreement with the experiment in the temperature range $J/k_{\rm B} \leq T  \leq 120$~K. Only at low temperatures the quantum approach yields more realistic results than the classical one. A big advantage of the classical approach is the fact that we can use convenient analytical expressions for the spin-spin correlation functions entering the second and fourth moments determining the ESR linewidth.

\acknowledgments

We thank Roghayeh Hassan Abadi and Mamoun Hemmida for experimental support. We are very grateful to Alexander I. Smirnov for useful comments. We acknowledge financial support by the Deutsche Forschungsgemeinschaft (DFG) via the Transregional Research Center TRR\,80 (Augsburg M\"unchen) and SFB/TR\,49.

\appendix

\section{}

Here we consider the contribution to the linewidth due to the anisotropic Zeeman interaction.
The Hamiltoinian for Cu$^{2+}$ (3d$^9$)-states is written as
\begin{equation}
{\cal H} = \sum_{k,q} B^{(k)}_{q}C^{(k)}_{q}+\lambda(\mathbf{S}\cdot\mathbf{L}). \label{CuHamiltonian}
\end{equation}
The first term describes the crystal-field effect and the second one the spin-orbit coupling. In a superposition model (or in other words in a model of independent metal-ligand bonds) the crystal-field parameters
can be calculated as follows
\begin{equation}
B^{(k)}_{q}=\sum_{j}a^{k}(R_{j})(-1)^{q}C^{(k)}_{q}(\theta_{j},\phi_{j}),  \label{Crystfield}
\end{equation}
where sum is performed over all surrounding ligands of the crystal lattice, and $C^{(k)}_{q}=\sqrt{\frac{4\pi}{2k+1}}Y^{k}_q$
are the components of the spherical tensor. According to Refs.~\onlinecite{Malkin1987, Eremin1977, Eremin1990} the so called intrinsic crystal-field parameters $a^{k}(R_{j})$ can be evaluated as
\begin{eqnarray} \label{crystfp}
\begin{split}
&a^{2}(R_{j})=-Z_{j}e^{2}\frac{\langle r^{2} \rangle_{3d}}{R^{3}_{j}}+G\frac{e^2}{R_j}(S^{2}_{3d\sigma}+S^{2}_{3d\pi}+S^{2}_{3ds}), \\
&a^{4}(R_{j})=-Z_{j}e^{2}\frac{\langle r^{4} \rangle_{3d}}{R^{5}_{j}}+G\frac{9e^2}{5R_j}(S^{2}_{3d\sigma}-\frac{4}{3}S^{2}_{3d\pi}+S^{2}_{3ds}).
\end{split}
\end{eqnarray}
Here $Z_{j}$ are the effective charges of the lattice ions, i.e. -1 for the Cl$^-$-ion, +1 for Cs$^+$ and +2 for Cu$^{2+}$.   $S_{3d\sigma}$, $S_{3d\pi}$, $S_{3ds}$ denote the overlap integrals between the Cu($3d^9$)-states and surrounding  $p_\sigma$, $p_\pi$ and $s$-orbitals of the Cl$^-$-ions. The overlap integrals in Eq.~\ref{crystfp} are calculated in local coordinate systems with the $z$-axis along the Cu \textendash Cl bond. In the present work these integrals were calculated using the Hartree \textendash Fock wave functions.\cite{Clementi1964, Clementi1974} The parameter $G$ was set to 14. Using this one adjustable parameter and the typical values of $\lambda=630$~cm$^{-1}$ and of the orbital momentum reduction factor $\kappa=0.8$ (due to covalency effect), we have calculated the energy level scheme, which is in reasonable agreement with experimental data \cite{Sharnov} (see Table~\ref{tabEnergy}).
\begin{table}[h]
\caption{\label{tabEnergy}Energy levels of Cu$^{2+}$(3d$^9$) in Cs$_2$CuCl$_4$ (cm$^{-1}$).}
\begin{ruledtabular}
\begin{tabular}{lc lc}
Calculated & Experiment \cite{Ferguson1964}\\
\hline
0 & 0\\
5060 & 4800\\
6260 & 5550\\
7110 & 7900\\
9026 & 9050
\end{tabular}
\end{ruledtabular}
\end{table}

Using the energy levels and corresponding wave functions we have calculated the components of the $g$-tensors for both magnetically nonequivalent positions of Cu in the unit cell of Cs$_2$CuCl$_4$ (see Fig.1 in Ref.~\onlinecite{Sharnov}).

For the ground-state Kramers doublet $|\pm 1/2 \rangle$ the obtained components ($\mathbf{g}_1$ -- for Cu(1) and Cu(3), $\mathbf{g}_2$ -- for Cu(2) and Cu(4)) read:
\begin{eqnarray}
\begin{split}
& \mathbf{g_1}=\begin{pmatrix} 2.260 & 0 & 0.402 \\ 0 & -2.088 & 0 \\ -0.056 & 0 & -2.453 \end{pmatrix}, \\[6px]
& \mathbf{g_2} = \begin{pmatrix} 2.260 & 0 & -0.403 \\ 0 & -2.086 & 0 \\ 0.056 & 0 & -2.453.  \end{pmatrix}.
\end{split}
\end{eqnarray}
Here we used the global coordinate system like it was chosen for description of the crystal structure.\cite{Krueger2010}
The corresponding effective spin Hamiltonian is defined as
\begin{equation}
{\cal H_{\rm Z}} = \mu_{\rm B} \cdot \sum_{\alpha,\beta} g_{\alpha \beta} H_{\alpha} S_{\beta}. \label{ZeemanHamiltonian}
\end{equation}

As one can see, the differences between the $g$-tensor components of these two Cu complexes are small, when the external magnetic field is applied along one of the main crystallographic axes. Therefore the anisotropic Zeeman contribution to the ESR linewidth
\begin{equation}
\Delta H_{\rm AZ} = \left(\frac{g_1-g_2}{g}\right)^2 \frac{g \mu_{\rm B} H_{\rm res}^2}{\sqrt{\langle J^2 \rangle}} \label{DHAZ}
\end{equation}
is expected to be small at least with respect to the DM contribution when considering the temperature dependence for the magnetic field applied along one of the main crystallographic axes. Here $\langle J^2 \rangle = (J^2+2(J')^2)/3$ denotes the averaged exchange integral as introduced in Ref.~\onlinecite{Huber1999}.

In order to be sure we have further checked this conclusion experimentally by measurements of the ESR linewidth at Q-band frequencies ($\approx$35~GHz) at $T=100$~K. At the same time one sees that the difference of the non diagonal components  $(g_{ac}^{(1)}-g_{ac}^{(2)})=+0.8$ is quite large. Therefore the additional broadening effect, which is quadratic in the applied
magnetic field, should be visible as soon as one changes the orientation of the field away from the main crystallographic directions. Indeed, our Q-band measurements have confirmed this effect. For fitting the experimental data in the upper panel of Fig.~4 the calculated $g$-components were slightly tuned as follows: $g_{aa}^{(1)}=g_{aa}^{(2)}=2.2$, $g_{bb}^{(1)}=g_{bb}^{(2)}=-2.08$, $g_{cc}^{(1)}=g_{cc}^{(2)}=-2.3$, $g_{ac}^{(1)}=-g_{ac}^{(2)}=0.25$, and $g_{ca}^{(1)}=-g_{ca}^{(2)}=-0.056$.

\section{}
Using the classical-spin model -- as has been done by Fisher\cite{Fisher1964} -- the temperature dependence of the pair-spin and  $\langle S^+ S^- \rangle$ correlation functions reads
\begin{eqnarray}
C_{n} = \frac{S(S+1)}{3}u^n, \label{Cn}
\end{eqnarray}
\begin{eqnarray}
\frac{\langle S^+ S^- \rangle}{N}=\frac{2}{3}S(S+1)\frac{1+u}{1-u}, \label{S+S-}
\end{eqnarray}
where $u = \coth K -1/K$ and  $K=-JS(S+1)/k_{\rm B}T$.

The four-spin contribution $C_{(4)}$ introduced in Eq.~\ref{M4a} -- in general case for a spin $S=1/2$ chain  -- after some  manipulations, can be written as follows
\begin{eqnarray*}
\langle  S^z_{i} S^z_{i+1}( S^z_{i+2}S^z_{i+3}- S^z_{i+3}S^z_{i+4}) \rangle +3\langle  S^y_{i} S^z_{i+1} S^y_{i+2}S^z_{i+3} \rangle
 \\ - \left[\langle S^y_{i} S^y_{i+1} S^x_{i+3}S^x_{i+4} \rangle+\langle S^y_{i} S^x_{i+1} (S^x_{i+2}S^y_{i+3}+S^x_{i+3}S^y_{i+4})\rangle\right].
\end{eqnarray*}
Using the classical spin fluctuation picture,\cite{Fisher1964} Nagata and Tazuke\cite{Nagata1972a} have derived the following analytical expressions:
\begin{eqnarray*}
\langle  S^\mu_{m} S^\mu_{m+1}S^\mu_{i}S^\mu_{j} \rangle &=& \frac{[S(S+1)]^2}{9}u^{j-i+1}\left[1+\frac{4}{5}v^{i-m-1}\right], \nonumber
\\ \langle  S^\mu_{m} S^\mu_{m+1}S^\gamma_{i}S^\gamma_{j} \rangle &=& \frac{\left[S(S+1)\right]^2}{9}u^{j-i+1}\left[1-\frac{2}{5}v^{i-m-1}\right]
\end{eqnarray*}
for $m+1<i<j$. With these expressions and the following one derived in the present work
\begin{eqnarray}
\langle  S^\mu_{m} S^\mu_{k}S^\gamma_{i}S^\gamma_{j} \rangle = \frac{\left[S(S+1)\right]^2}{15}u^{j-i+m-k}v^{k-m}
\end{eqnarray}
for $i<m<k$, $j>k$,
\begin{eqnarray}
\langle  S^\mu_{m} S^\gamma_{m+1}S^\mu_{i}S^\gamma_{i+1} \rangle &=& \langle  S^\mu_{m} S^\gamma_{m+1}S^\gamma_{i}S^\mu_{i+1} \rangle \nonumber \\ &=& \frac{\left[S(S+1)\right]^{2}}{15}u^{2}v^{i-m-1} \label{cor}
\end{eqnarray}
for $i>m+1$,  $\mu,\gamma=(x,y,z)$  and  $v=1-3u/K$, finally we arrived at
\begin{eqnarray}
C_{(4)} = -\frac{\left[S(S+1)\right]^2}{9}u^2(v-1)^2.
\end{eqnarray}
It should be noted that during the derivation of the general expression in the second moment, the four-spin correlation functions also appear in the following combination
\begin{eqnarray}
\sum_{|j-m|\geq2}\langle  S^\mu_{m} S^\gamma_{m+1}(S^\mu_{j}S^\gamma_{j+1}-S^\gamma_{i}S^\mu_{j+1}) \rangle.
\end{eqnarray}
However, in the quasi classical approach with the use of Relation~\ref{cor}, it becomes clear that this contribution vanishes and, hence, it was not considered in Eq.~\ref{M4a}


%

\end{document}